\documentclass[aps,prb,twocolumn,epsf,float,letter]{revtex4}

\usepackage{graphicx}
\usepackage{bm}
\usepackage{setspace}
\usepackage{amsmath}
\usepackage{braket}
\begin{document}
\title{Angular dependence of spin-orbit spin transfer torques}

\author{Ki-Seung Lee$^{1,**}$}
\author{Dongwook Go$^{2,**}$}
\author{Aur\'{e}lien Manchon$^3$}
\author{Paul M. Haney$^4$}
\author{M. D. Stiles$^4$}
\author{Hyun-Woo Lee$^2$}
\email{hwl@postech.ac.kr}
\author{Kyung-Jin Lee$^{1,5}$}
\email{kj_lee@korea.ac.kr}

\affiliation{$^1$Department of Materials Science and Engineering, Korea University, Seoul 136-701, Korea \\$^2$PCTP and Department of Physics, Pohang University of Science and Technology, Pohang 790-784, Korea \\$^3$Physical Science and Engineering Division, King Abdullah University of Science and Technology (KAUST), Thuwal 23955-6900, Saudi Arabia \\$^4$Center for Nanoscale Science and Technology, National Institute of Standards and Technology, Gaithersburg, Maryland 20899, USA \\$^5$KU-KIST Graduate School of Converging Science and Technology, Korea University, Seoul 136-713, Korea}

\begin{abstract}
  In ferromagnet/heavy metal bilayers, an in-plane current gives rise to spin-orbit spin transfer torque which is usually decomposed into field-like and damping-like torques. For two-dimensional free-electron and tight-binding models with Rashba spin-orbit coupling, the field-like torque acquires nontrivial dependence on the magnetization direction when the Rashba spin-orbit coupling becomes comparable to the exchange interaction. This nontrivial angular dependence of the field-like torque is related to the Fermi surface distortion, determined by the ratio of the Rashba spin-orbit coupling to the exchange interaction. On the other hand, the damping-like torque acquires nontrivial angular dependence when the Rashba spin-orbit coupling is comparable to or stronger than the exchange interaction. It is related to the combined effects of the Fermi surface distortion and the Fermi sea contribution. The angular dependence is consistent with experimental observations and can be important to understand magnetization dynamics induced by spin-orbit spin transfer torques.
\end{abstract}

 \maketitle

\section{Introduction}
\label{sec:intro}

In-plane current-induced spin-orbit spin transfer torques in
ferromagnet/heavy metal bilayers provide an efficient way of inducing
magnetization dynamics and may play a role in future magnetoelectronic
devices.~\cite{Miron:2011a, Miron:2011b, Liu:2012a, Liu:2012b,
  Seo:2012, Thiaville:2012, Haazen:2013, Emori:2013, Ryu:2013,
  Lee:2013, Brink:2013, Garello:2013a, Lee:2014, Yoshimura:2014} Two
mechanisms for spin-orbit torques have been proposed to date; the bulk
spin Hall effect in the heavy metal layer~\cite{DP:1971, Hirsch:1999,
  Zhang:2000, Ando:2008, Liu:2011} and interfacial spin-orbit coupling
effect at the ferromagnet/heavy metal interface~\cite{Bychkov:1984,
  Edelstein:1990, Obata:2008, Manchon:2008, MatosAbiague:2009,
  Wang:2012, Kim:2012, Pesin:2012, Bijl:2012, Haney1:2013,
  Kurebayashi:2014} frequently referred to as the Rashba
effect. Substantial efforts have been expended in identifying the
dominant mechanism for the spin-orbit torque.~\cite{Haney1:2013,
  Kurebayashi:2014, KimJ:2013, Qiu:2013, Fan:2014, Liu:2014}  For this
purpose, one needs to go beyond qualitative analysis since both the
mechanisms result in qualitatively identical predictions, i.e. two
vector components of spin-orbit torque (see Eq.~(\ref{Eq:SOT})). For
quantitative analysis, we adopt the commonly used decomposition of the
spin-orbit torque ${\bf T}$,
\begin{equation}
  {\bf T} = \tau_{\rm f} {\hat {\bf M}} \times {\hat {\bf y}} + \tau_{\rm d}  {\hat {\bf M}} \times ({\hat {\bf M}} \times {\hat {\bf y}}),
  \label{Eq:SOT}
\end{equation}
where the first term is commonly called the field-like spin-orbit
torque, the second term the damping-like spin-orbit torque or the
Slonczewski-like spin-orbit torque, ${\hat {\bf M}} = (\cos \phi \sin
\theta, \sin \phi \sin \theta, \cos \theta)$ is the unit vector along
the magnetization direction, $ {\hat {\bf y}}$ is the unit vector
perpendicular to both current direction ($ {\hat {\bf x}}$) and the
direction in which the inversion symmetry is broken ($ {\hat {\bf
    z}}$), $\tau_{\rm f}$ and $\tau_{\rm d}$ describe the magnitude of
field-like and damping-like spin-orbit torque terms,
respectively. Since ${\bf T}$ should be orthogonal to ${\hat {\bf
    M}}$, the two terms in Eq.~(\ref{Eq:SOT}), which are orthogonal to
${\hat {\bf M}}$ and also to each other, provide a perfectly general
description of the spin-orbit torque regardless of the detailed
mechanism of ${\bf T}$. The quantitative analysis of ${\bf T}$ then
amounts to the examination of the properties of $\tau_{\rm f}$ and
$\tau_{\rm d}$.

One of intriguing features of spin-orbit torque observed in some
experiments is the strong dependence of $\tau_{\rm f}$ and $\tau_{\rm
  d}$ on the magnetization direction.~\cite{Garello:2013, Qiu:2014}
Comparing the measured and calculated angular dependence will provide clues to the
mechanism of the spin-orbit torque.  The detailed angular dependence
also determines the magnetization dynamics and hence is
important for device applications based on
magnetization switching~\cite{Miron:2011a, Liu:2012a, Lee:2013,
  Brink:2013, Garello:2013a, Lee:2014}, domain wall
dynamics~\cite{Miron:2011b, Seo:2012, Thiaville:2012, Haazen:2013,
  Emori:2013, Ryu:2013, Yoshimura:2014}, and magnetic skyrmion
motion~\cite{Sampaio:2013}.

Theories based on the bulk spin Hall effect combined with a
drift-diffusion model or Boltzmann transport
equation~\cite{Haney1:2013} predict no angular dependence of
$\tau_{\rm f}$ and $\tau_{\rm d}$, which is not consistent with the
experimental results.~\cite{Garello:2013, Qiu:2014} For theories based
on the interfacial spin-orbit
coupling, the angular dependence is subtle. Based on the Rashba model
including D'yakonov-Perel spin
relaxation, Pauyac {\it et al.}~\cite{Pauyac:2013} studied the angular
dependence of spin-orbit torque perturbatively in weak Rashba regime
($r \equiv \alpha_{\rm R} k_{\rm F}/J \ll 1$) and strong Rashba regime
($r \gg 1$) where $\alpha_{\rm R}$ is the strength of the Rashba
spin-orbit coupling, $k_{\rm F}$ is the Fermi wave vector, and $J$ is
the exchange coupling. They found that both $\tau_{\rm f}$ and
$\tau_{\rm d}$ are almost independent of the angular direction of
${\hat {\bf M}}$ in the weak Rashba regime. In the strong Rashba
regime, on the other hand, they found that $\tau_{\rm d}$ exhibits
strong angular dependence. The origin of the angular dependence within
this model is the anisotropy of the spin relaxation, which arises
naturally since the Rashba spin-orbit interaction is responsible for
the anisotropic D'yakonov-Perel spin relaxation mechanism. For
$\tau_{\rm f}$, in contrast, they found it to be almost constant in
the strong Rashba regime even when the spin relaxation is
anisotropic.  Experimentally,~\cite{Garello:2013, Qiu:2014} both the
damping-like and the field-like contributions depend strongly on the
magnetization direction.

Here we reexamine the angular dependence of the spin-orbit torque
based on the Rashba interaction motivated by the following two
observations. The first motivation comes from a first-principles
calculation~\cite{Haney2:2013} of Co/Pt bilayers, according to which
both the spin-orbit potential and the exchange splitting are large
near the interface between the heavy metal and the ferromagnet. This
implies that the problem of interest is not in the analytically
tractable weak Rashba or strong Rashba regime but in the intermediate
Rashba regime ($r \approx 1$). We examine this intermediate regime
numerically and find that in contrast to both the strong and weak
Rashba regimes, $\tau_{\rm f}$ has a strong angular dependence. The
second motivation comes from a recent calculation~\cite{Bijl:2012,
  Kurebayashi:2014} showing that the interfacial spin-orbit coupling
can generate $\tau_{\rm d}$ through a Berry phase
contribution~\cite{Sinova:2004}. In contrast, earlier
theories~\cite{Wang:2012, Kim:2012, Pesin:2012} of the interfacial
spin-orbit coupling found a separate contribution to $\tau_{\rm d}$
from spin relaxation. Moreover those
calculations~\cite{Kurebayashi:2014} suggest that the Berry phase
contribution to $\tau_{\rm d}$ is much larger than the spin relaxation
contribution. Here, we examine the angular dependence of the Berry
phase contribution.

To be specific, we examine the angular dependence of the spin-orbit
torques for a free-electron model of two-dimensional ferromagnetic
systems with the Rashba spin-orbit coupling. When an electric field is
applied to generate an in-plane current, the spin-orbit torque arises
from the two types of changes caused by the electric field. One is the
electron occupation change.  For a small electric field, the net
occupation change is limited to the Fermi surface so that the
spin-orbit torque caused by the occupation change comes from the Fermi
surface. For this reason, this contribution is referred to as the
Fermi surface contribution. The other is the state change. The
electric field modifies the potential energy of the system, which in
turn modifies wavefunctions of all single particle states. Thus the
spin-orbit torque caused by the state change comes not only from the
states near the Fermi surface but also from all states in the entire
Fermi sea. This contribution is referred to as the Fermi sea
contribution and often closely related to the momentum-space Berry
phase~\cite{Kurebayashi:2014}.

We find that in the absence of spin relaxation, the Fermi surface
contribution to $\tau_{\rm d}$ is vanishingly small, while $\tau_{\rm
  f}$ remains finite. The $\tau_{\rm f}$ has a substantial angular
dependence in the intermediate Rashba regime. This nontrivial angular
dependence of $\tau_{\rm f}$ is related to Fermi surface distortion,
which becomes significant when the Rashba spin-orbit coupling energy
($\sim \alpha_R k_F$) is comparable to the exchange coupling ($\sim
J$).  On the other hand, the Fermi sea contribution generates
primarily $\tau_{\rm d}$ which exhibits strong angular dependence in
both the intermediate and strong Rashba regimes. The nontrivial
angular dependence of $\tau_{\rm d}$ is caused by the combined effects
of the Fermi surface distortion and the Fermi sea contribution. We
also compute the angular dependence of the spin-orbit torques for a
tight-binding model and find that the results are qualitatively
consistent with those for a free-electron model.

\section{Semiclassical models}
\label{sec:model}

In this section, we use subscripts (1) and (2) to denote the Fermi
surface and the Fermi sea contributions, respectively. The model
Hamiltonian for an electron in the absence of an external electric
field is
\begin{equation}
  H_0 = \frac{{\bf p}^2}{2m} + \alpha_{\rm R}  {\boldsymbol{\sigma}} \cdot  (\bf{k}
  \times \bf{\hat{z}}) +
  {\it{J}}\boldsymbol{\sigma} \cdot \bf{\hat{M}},
  \label{Eq:Hfem}
\end{equation}
where ${\bf k}=(k_x, k_y)$ is the two-dimensional wave vector, $m$ is
the electron mass, $J$ ($>$0) is the exchange parameter, ${\bf p}$ is
the momentum, and ${\boldsymbol \sigma}$ is the vector of Pauli
matrices, and $M_x$, $M_y$, and $M_z$ are the $x$-, $y$-, and
$z$-components of $\hat{\bf M}$, respectively. When ${\hat {\bf M}}$
is position-independent, which will be assumed all throughout this
paper, ${\bf k}$ is a good quantum number. For each ${\bf k}$, there
are two energy eigenvalues since the spin may point in two different
directions. Thus the energy eigenvalues of $H_0$ form two energy
bands, called majority and minority bands. The one-electron
eigenenergy of $H_0$ is
\begin{equation}
  E_{\bf{k}}^{\pm} = \frac{\hbar^2 k^2}{2m} \mp \epsilon_{\bf{k}} ,
  \label{Eq:Efem}
\end{equation}
where the upper (lower) sign corresponds to the majority (minority) band, $k^2 =k_x^2+k_y^2$, and $\epsilon_{\bf{k}} = | J
\bf{\hat{M}}+ \alpha_{\rm R} (\bf{k} \times \bf{\hat{z}}) |$.

To determine the spin state of the majority and minority bands, it is
useful to combine the last two terms of $H_0$ into an effective Zeeman
energy term ($= -\mu_B {\bf B}_{\rm eff, \bf k} \cdot {\boldsymbol
  \sigma}$), where the effective magnetic field is ${\bf k}$-dependent
and given by
\begin{equation}
  {\bf{B}}_{{\rm eff},\bf{k}} = - \frac{J}{\mu_{\rm B}} \bf{\hat{M}} -
  \frac{\alpha_{\rm R}}{\mu_{\rm B}} ({\bf{k}} \times \bf{\hat{z}}).
  \label{Eq:Bfem}
\end{equation}
Here $\mu_{\rm B}$ is the Bohr magneton. ${\bf B}_{\rm eff, \bf k}$
fixes the spin direction of the majority and minority bands. For the
eigenstate $\ket {\psi_{{\bf k},\pm}}$ of an eigenstate in the
majority/minority band, its spin expectation value ${\bf s}^\pm_{{\bf
    k}(1)} \equiv (\hbar/2) \bra{\psi_{{\bf k},\pm}} {\boldsymbol
  \sigma} \ket{\psi_{{\bf k},\pm}}$ is given by
\begin{equation}
  {{\bf s}_{{\bf k}(1)}^\pm} =  \pm \frac{\hbar}{2} {\bf{\hat
      B}}_{{\rm eff},\bf{k}},
  \label{Eq:skfem1}
\end{equation}
where ${\bf{\hat B}}_{{\rm eff},{\bf k}}$ is the unit vector along
${\bf{B}}_{{\rm eff},{\bf k}}$. In terms of the ${\bf k}$-dependent
angle $\theta_{\bf k}$ and $\phi_{\bf k}$, which are defined by
$\hat{\bf B}_{\rm eff, \bf k} = (\sin \theta_{\bf k} \cos \phi_{\bf
  k}, \sin \theta_{\bf k} \sin \phi_{\bf k}, \cos \theta_{\bf k})$,
the eigenstate $\ket{\psi_{{\bf k},\pm}}$ is given by
\begin{eqnarray}
  \ket{\psi_{{\bf k},+}} &=& {\rm e}^{i {\bf k} \cdot {\bf r}} \begin{pmatrix} \cos (\theta_{\bf k}/2) \\ \sin (\theta_{\bf k}/2) {\rm e}^{i \phi_{\bf k}}  \end{pmatrix} \\
  \ket{\psi_{{\bf k},-}} &=& {\rm e}^{i {\bf k} \cdot {\bf r}} \begin{pmatrix} \sin (\theta_{\bf k}/2) \\ -\cos (\theta_{\bf k}/2) {\rm e}^{i \phi_{\bf k}}  \end{pmatrix}
\end{eqnarray}
Together with the energy eigenvalue $E^\pm_{\bf k}$ in
Eq.~(\ref{Eq:Efem}), the eigenstate $\ket{\psi_{{\bf k},\pm}}$
completely specifies properties of the equilibrium Hamiltonian. The
ground state of the system is then achieved by filling up all single
particle eigenstates $\ket{\psi_{{\bf k},\pm}}$, below the Fermi
energy $E_F$.

When an electric field ${\bf E}=E \bf{\hat{x}}$ is applied, one of the
effects is the modification of the state
occupation. This effect generates the non-equilibrium spin density
${\bf s}^\pm_{(1)}$ as
\begin{equation}
  {\bf s}_{(1)}^\pm = \int \frac{dk^2}{(2\pi)^2}  \left[ f_{\pm} \left(
      {\bf{k}}-\frac{eE \tau}{\hbar} {\hat{\bf x}} \right) - f_{\pm}
    \left( {\bf{k}} \right) \right]   {{\bf s}_{{\bf k}(1)}^\pm},
  \label{Eq:sdfem1}
\end{equation}
where $-e$ is the electron charge, $\tau$ is the relaxation time, and
$f_{\pm} ({\bf k})= \Theta (E_F-E_{\bf k}^\pm) $ is the
zero-temperature electron
occupation function where $\Theta(x)$ is the Heaviside step function.
Note that the net contribution to ${\bf s}^\pm_{(1)}$ arises entirely
from the states near $E_F$ due to the cancellation effect between the
two occupation functions in Eq.~(\ref{Eq:sdfem1}). Thus ${\bf
  s}^\pm_{(1)}$ is a {\it Fermi surface}
contribution. The total spin density generated by the occupation
change becomes ${\bf s}_{(1)}={\bf s}^{+}_{(1)}+{\bf
  s}^{-}_{(1)}$. This is related to the spin-orbit torque ${\bf
  T}_{(1)}$ generated by the occupation change via ${\bf
  T}_{(1)}=(J/\hbar) {\bf s}_{(1)} \times \hat{\bf M}$. In
Eq.~(\ref{Eq:sdfem1}), we use the relaxation time approximation with
the assumption that the scattering probability is isotropic and
spin-independent.

The other important effect of the electric field other changing than
the occupation is that it modifies the potential
energy that the electrons feel, and hence modifies their
wave functions, generating in turn a correction to ${\bf s}^\pm_{{\bf
    k}(1)}$. We
call this correction ${\bf s}^\pm_{{\bf k}(2)}$. It is calculated in
Appendix~\ref{sec:appendixA} and given by
\begin{equation}
  {{\bf s}_{{\bf k}(2)}^\pm}= \pm\frac{\hbar}{2}\alpha_R e E \left[
    \frac{ J}{2\epsilon_\mathbf{k}^3}(\hat{\mathbf{M}}\times \hat{\mathbf{y}})+\frac{\alpha_R}{2\epsilon_\mathbf{k}^3}(\hat{\mathbf{x}}\times \mathbf{k})\right].\label{Eq:skfem2}
\end{equation}
Summing over all occupied states in the majority/minority band,
gives the total spin density ${\bf s}^\pm_{(2)}$ generated by
the state change in that band, and is given by
\begin{equation}
  {\bf s}_{(2)}^\pm = \int \frac{dk^2}{(2\pi)^2}  f_{\pm} ({\bf{k}})   {{\bf s}_{{\bf k}(2)}^\pm}.
  \label{Eq:sdfem2}
\end{equation}
Note that the equilibrium occupation function $f$ appears in
Eq.~(\ref{Eq:sdfem2}) rather than the difference between the two
occupation functions. The occupation change effect is ignored in
Eq.~(\ref{Eq:sdfem2}) since we are interested in linear effects of the
electric field $E$ and ${\bf s}^\pm_{{\bf k}(2)}$ is already first
order in $E$. Note that all the occupied single particle states in the Fermi
sea contribute to ${\bf s}^\pm_{(2)}$. Thus ${\bf s}^\pm_{(2)}$
amounts is a {\it Fermi sea} contribution. The total spin
density generated by the state change becomes ${\bf s}_{(2)}={\bf
  s}^{+}_{(2)}+{\bf s}^{-}_{(2)}$. This is related to the spin-orbit
torque ${\bf T}_{(2)}$ generated by the state change via ${\bf
  T}_{(2)}=(J/\hbar) {\bf s}_{(2)} \times \hat{\bf M}$.

A few remarks are in order. In Eq.~(\ref{Eq:sdfem1}), the two
occupation functions cancel each other for most ${\bf k}$ values. They
do not
cancel for ${\bf k}$ points that correspond to
electron excitation slightly above the Fermi surface or the hole
excitation slightly below the Fermi surface. Thus the direction of
${\bf s}^\pm_{(1)}$ can be estimated simply by evaluating the
difference of ${\bf B}_{\rm eff,{\bf k}}$ between two ${\bf k}$'s of
electron-like and hole-like excitations. This shows that ${\bf
  s}^\pm_{(1)}$ points along $E \hat{\bf x} \times \hat{\bf z} = - E
\hat{\bf y}$. Thus the spin-orbit torque ${\bf T}_{(1)}$ should be
proportional to $E \hat{\bf y} \times \hat{\bf M}$, which is nothing
but the field-like spin-orbit torque. Thus the Fermi surface
contribution ${\bf T}_{(1)}$ contributes mostly to $\tau_{\rm f}$. To be
precise, however, this statement is not valid for
spin-dependent scattering, which we neglect in deriving
Eq.~(\ref{Eq:sdfem1}). If the scattering is
spin-dependent, ${\bf T}_{(1)}$ produces $\tau_{\rm d}$ as well as
$\tau_{\rm f}$ as demonstrated in Refs.~\onlinecite{Wang:2012,
  Kim:2012, Pesin:2012}. In this paper, we
neglect the contribution to the angular dependence of $\tau_{\rm d}$
from ${\bf T}_{(1)}$ and
spin-dependent scattering since it has been already treated in
Ref.~\onlinecite{Pauyac:2013}. The contribution to $\tau_{\rm d}$ in our study comes from the Fermi
sea contribution ${\bf T}_{(2)}$. One can easily verify that the first
term in Eq.~(\ref{Eq:skfem2}) generates the spin-orbit torque
proportional to $(\hat{\bf M} \times \hat{\bf y}) \times \hat{\bf M}$,
which has the form of the damping-like spin-orbit torque. The second
term in Eq.~(\ref{Eq:skfem2}) on the other hand almost vanishes upon
${\bf k}$ integration in Eq.~(\ref{Eq:sdfem2}). This demonstrates
that the Fermi sea contribution ${\bf T}_{(2)}$ contributes mostly to
$\tau_{\rm d}$.

\begin{figure*}
  \includegraphics[width=0.9\textwidth]{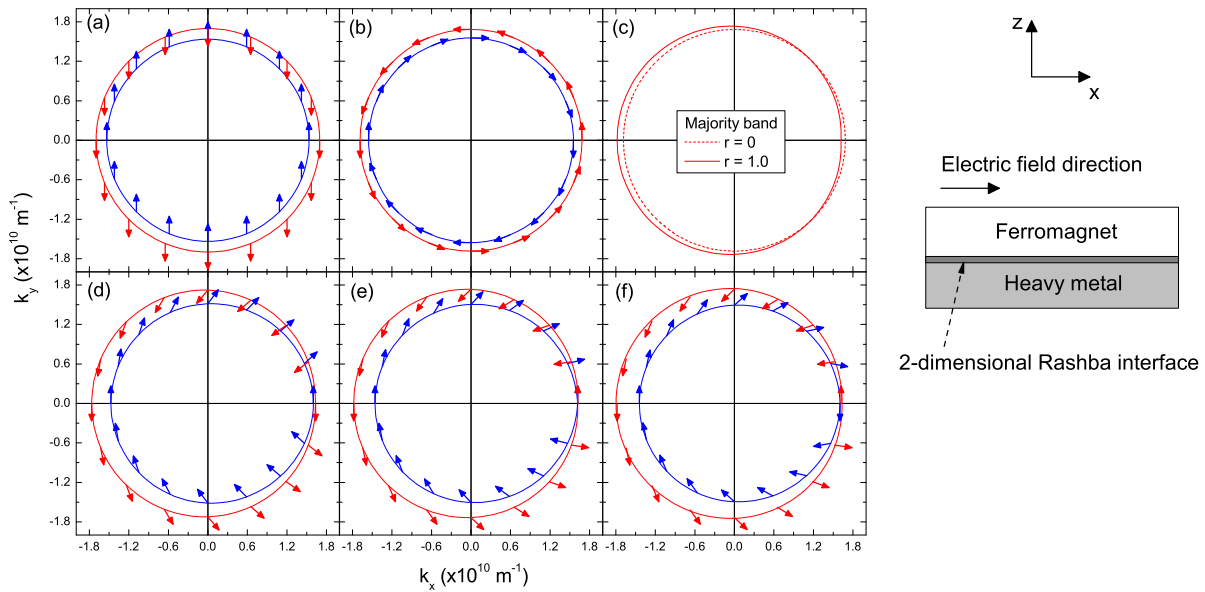}
  \caption{(color online) Fermi surface and spin direction for a free-electron model. (a) $r=0$ (only exchange splitting), (b) $r= \infty $ (only Rashba spin-orbit coupling and non-magnetic), (c) comparison of Fermi surfaces (majority band) for $r=0$ and $r=1.0$, (d) $r=0.8$, (e) $r=1.0$, and (f) $r=1.2$. We assume ${\bf{\hat{M}}} = (0, 1, 0)$, $E_{\rm F}$= 10 eV, $m=m_0$, and $J$= 1 eV. Here $m_0$ is the free electron mass. The outer red (inner blue) Fermi surface corresponds to majority (minority) band. Arrows are the eigendirections of spins on the Fermi surface. The coordinate system is shown on the right.} \label{Fig1}
\end{figure*}

We also compute the spin-orbit torques based on a tight-binding model
because free electron models with linear Rashba coupling, like we use
here, can exhibit pathological behavior when accounting for vertex
corrections to the impurity scattering.  For example, the intrinsic
spin Hall effect~\cite{Sinova:2004}, that has the same physical origin
of the Fermi sea contribution to spin-orbit torque, gives a universal
result that vanishes when vertex corrections are
included.~\cite{Inoue:2003, Schliemann:2004, Burkov:2004,
  Mishchenko:2004, Dimitrova:2005} However, the intrinsic spin Hall
effect does not vanish when the electron dispersion deviates from free
electron behavior or the spin-orbit coupling is not linear in
momentum.~\cite{Murakami:2004, Bernevig:2005, Shytov:2006,
  Khaetskii:2006} Since we neglect vertex corrections in the
calculations presented in this paper, it is necessary to check whether
or not the angular dependence of spin-orbit torque obtained in a
free-electron model is qualitatively reproduced in a tight-binding
model, where the electron dispersion deviates from free electron
behavior and the spin-orbit coupling is not strictly linear in
momentum.  To compute spin-orbit torque in the two-band (majority and
minority spin bands) tight-binding model on a square lattice with the
lattice constant $a$, we replace $k_x$ and $k_y$ by $\sin (k_x a)/a$
and $\sin (k_y a)/a$, respectively. The corresponding spin density is
then calculated by integrating the electric field-induced spin
expectation value up to the point of band filling. For most cases in a
tight-binding model, the result converges for a ${\bf k}$-point mesh
with mesh spacing $dk$=0.052 nm$^{-1}$ and 80,000 ${\bf k}$-points,
where the convergence criteria is 1 percent change of results with a
finer mesh by factor of 2.  All results presented in this paper are
converged to this criteria.

\section{Results and Discussion}
\label{sec:result1}

We first discuss Fermi surface distortion as a function of $r (=
\alpha_{\rm R} k_{\rm F}/J)$. Figure \ref{Fig1} shows the Fermi
surface and the spin direction at each ${\bf k}$-point for various
values of
the ratio $r$.  Without Rashba spin-orbit coupling $(r=0)$, the spin
direction does not depend on $\bf k$ for ferromagnetic systems
(Fig. \ref{Fig1}(a)).  Without exchange coupling (non-magnetic Rashba
system $(r=\infty)$), on the other hand, the spins point in the
azimuthal direction (Fig. \ref{Fig1}(b)). For these extreme cases, the
Fermi surfaces of two bands are concentric circles.

The Fermi surfaces distort significantly when $r \approx
1$. Figure \ref{Fig1}(c) compares two Fermi surfaces (majority band)
for $r=0$ and $r=1.0$ when ${\bf{\hat{M}}} = (0, 1, 0)$.   When the
magnetization has an in-plane component as in this case, each sheet of
the Fermi
surface shifts in a different direction and distorts
from perfect circularity (Fig. \ref{Fig1}(c): note that the
dotted Fermi surface is for $r=0$ and is a circle). This
distortion arises because the ${\bf k}$-dependent effective magnetic
field (Eq.~(\ref{Eq:Bfem})) contains contributions both from the
exchange and Rashba spin-orbit couplings. An effective field from the
exchange is aligned along ${\bf{\hat{M}}}$ and uniform regardless of
${\bf k}$, whereas that from the Rashba spin-orbit coupling lies in
the $x-y$ plane and is ${\bf k}$-dependent.  For example, for
${\bf{\hat{M}}} = (0, 1, 0)$ and majority band, an effective field
from the Rashba spin-orbit coupling is parallel (anti-parallel) to
that from the exchange at ${\bf k}=(k_{F,1},0)$ (${\bf
  k}=(k_{F,2},0)$), where $k_{F,1}$ ($>$ 0) and $k_{F,2}$ ($<$ 0) are
the Fermi wave vectors corresponding to the electric field-induced
electron-like and hole-like excitations, respectively. The ${\bf
  k}$-dependent effective field distorts
the Fermi surface distortion as demonstrated in
Fig. \ref{Fig1}(c)-(f).

This Fermi surface distortion also affects the spin direction at each
${\bf k}$-point because the spin eigendirection is ${\bf k}$-dependent
due to the Rashba spin-orbit coupling. In the weak (strong) Rashba
regime, the spin landscape is similar with that in Fig. \ref{Fig1}(a)
(Fig. \ref{Fig1}(b)). In these extreme cases, the spin landscape is
not significantly modified by the change in the magnetization
direction as one of the effective fields (either from the exchange or
from the Rashba spin-orbit coupling) is much stronger than the
other. As a result,  $\tau_{\rm f}$ has almost no angular distortion in these
regimes.  The spin landscape for $r \approx 1$ on
the other hand becomes highly complicated (Fig. \ref{Fig1}(d)-(f)) as
the Fermi surface distortion is maximized. One can
easily verify that the spin landscape for $r \approx 1$ varies
significantly with the magnetization direction because the Fermi
surface distortion is closely related to the in-plane component of the
magnetization as explained above.

As the non-equilibrium spin density corresponding to $\tau_{\rm f}$
(i.e. ${\bf s}_{(1)}$) is obtained from the integration of the spins
on the Fermi surface, this magnetization-angle-dependent change in the
spin landscape generates a nontrivial angular dependence of
$\tau_{\rm f}$.  A similar argument is valid for $\tau_{\rm d}$
(i.e. ${\bf s}_{(2)}$) which comes from the Fermi sea contribution
because the Fermi surface distortion affects the interval of the
integration. Therefore, the results shown in Fig. \ref{Fig1} suggest
that the spin-orbit torque originating from the interfacial spin-orbit
coupling should have a strong dependence on the magnetization angles
$\theta$ and $\phi$ when $r$ is close to 1.

Several additional remarks for the Fermi surface distortion are as
follows. First, the two Fermi surfaces touch exactly for $r=1$
(Fig. \ref{Fig1}(e)) and they anticross for $r>1$
(Fig. \ref{Fig1}(f)). As a result, the spin landscape rapidly changes
when $r$ varies around 1 so that a similar drastic change in the
angular dependence of the spin-orbit torque is expected. Second, all
effects from the Fermi surface distortion, described for a
free-electron model above, should also affect the results obtained for
a tight-binding model. However, as the shape of the Fermi surface is
different for the two models (i.e. for $J=0$ and $\alpha_R=0$, the
Fermi surface for a free-electron model is a circle, whereas that for
a tight-binding model with half band-filling is a rhombus), the
results for the two models are quantitatively different.

\begin{figure}
  \includegraphics[width=1.0\columnwidth]{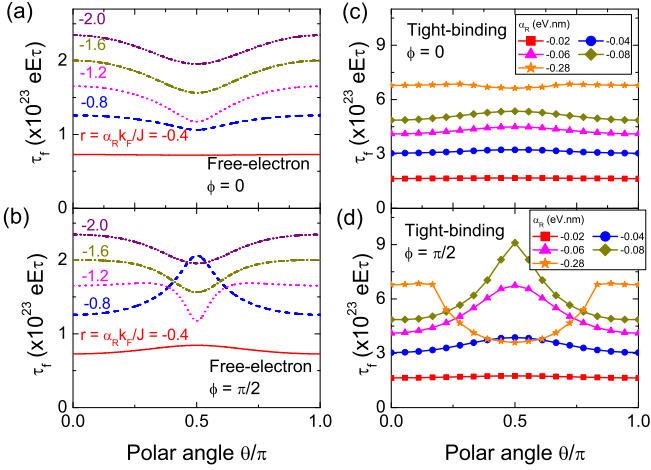}
  \caption{(color online) Polar angle ($\theta$) dependence of field-like spin-orbit torque coefficient $\tau_{\rm f}$. Free-electron model (a and b): (a) azimuthal angle of magnetization $\phi$ = 0 and (b) $\phi=\pi/2$. Tight-binding model (c and d): (c) $\phi=0$ and (d) $\phi=\pi/2$. For a free-electron model, we use $E_F$= 10 eV, $m=m_0$, and $J$= 1 eV. For a tight-binding model, we use $m=m_0$, $J$= 1 eV, $a=0.3$ nm, and normalized electron density $n= N/N_{\rm max} = 0.5$ where $N$ is the electron density of filled bands and $N_{\rm max}$ (= 2.2 $\times$ 10$^{19}$ m$^{-2}$) is the maximum electron density. For a free-electron model, the results for $r=1$ are excluded because it is singular. For a tight-binding model, the results for -0.28 eV$\cdot$nm $< \alpha_R <$ -0.08 eV$\cdot$nm are not included because of bad convergence. } \label{Fig2}
\end{figure}

We next show the angular dependence of $\tau_{\rm
  f}$ and $\tau_{\rm d}$ for the two models. Here we do not attempt to
analyze the detailed angular dependence quantitatively, because it is
very parameter sensitive. In contrast, our intention is to identify
the general trends that emerge from these numerical
calculations. Figure \ref{Fig2} shows the angular dependence of
$\tau_{\rm f}$ for a free-electron model ((a) and (b)) and a
tight-binding model ((c) and (d)). In both models, we obtain
nontrivial angular dependence of $\tau_{\rm f}$ in certain parameter
regimes. In the free-electron model, we find $\tau_{\rm f}$ is almost
constant in weak ($r \ll 1$) and strong ($r \gg 1$) Rashba regimes,
consistent with earlier works.\cite{Obata:2008, Manchon:2008} In the
intermediate Rashba regimes, however, $\tau_{\rm f}$ is not a
constant. We find that $\tau_{\rm f}$ depends not only on the polar
angle $\theta$ but also the azimuthal angle $\phi$, as expected from
the Fermi surface distortion (Fig. \ref{Fig1}). In Fig. \ref{Fig2}(b),
$\tau_{\rm f}$ for $r<1$ ($r>1$) is maximal (minimal)
at $\theta=\pi/2$, which is caused by the anticrossing of the two
Fermi surfaces (Fig. \ref{Fig1}(d)-(f)). Despite the strong angular
dependence, the sign of $\tau_{\rm f}$ is preserved since the
spin direction of nonequilibrium spin density is unambiguously
determined once the direction of electric field and the sign of
$\alpha_{\rm R}$ are fixed. These overall trends are qualitatively
reproduced in a tight-binding model (Fig. \ref{Fig2}(c) and (d)). The
magnitude and angular dependence of $\tau_{\rm f}$ differ
quantitatively from those of the free electron model, due to the
different shape of the Fermi surfaces for the two models.

\begin{figure}
  \includegraphics[width=1.0\columnwidth]{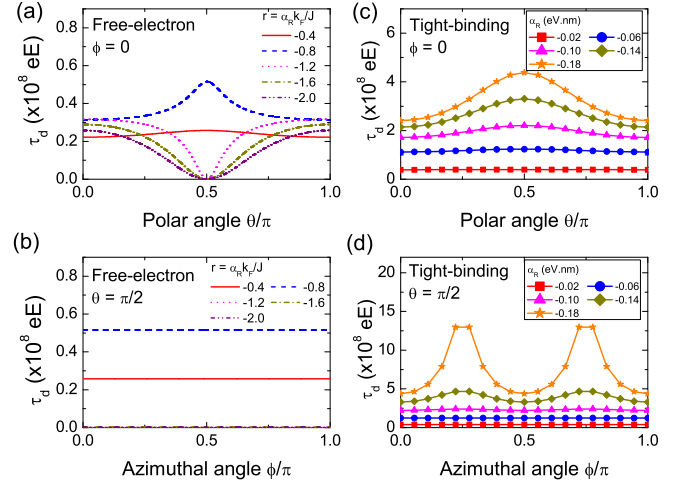}
  \caption{(color online) Angular dependence of damping-like
    spin-orbit torque coefficient $\tau_{\rm d}$. Free-electron model
    (a and b): (a) polar angle dependence at $\phi=0$ and (b)
    azimuthal angle dependence at $\theta=\pi/2$. Tight-binding model
    (c and d): (c) polar angle dependence at $\phi=0$ and (d)
    azimuthal angle dependence at $\theta=\pi/2$. Same parameters are
    used as in Fig. \ref{Fig2}. } \label{Fig3}
\end{figure}

Figure \ref{Fig3} shows the angular dependence of $\tau_{\rm d}$ for a
free-electron model ((a) and (b)) and a tight-binding model ((c) and
(d)). This $\tau_{\rm d}$ results from the Fermi sea contribution
(Eqs. (\ref{Eq:skfem2}) and (\ref{Eq:sdfem2})). In both models, we
obtain nontrivial angular dependence of $\tau_{\rm d}$ both in the
intermediate and strong Rashba regimes (Fig.~\ref{Fig3}(a) and
(c)). This is in contrast to $\tau_{\rm f}$ which exhibits nontrivial
angular dependence only in the intermediate Rashba regime.  To
understand this difference, we derive an approximate $\tau_{\rm d}$
by expanding up to third order in $\frac{\alpha_R k_F}{J}$ and
assuming no Fermi surface distortion (i.e. the Fermi wave vector $k_F$
does not depend on the direction of $\bf k$), which is analytically
tractable. By integrating Eq. (\ref{Eq:sdfem2}) with these
assumptions, we find $\tau_{\rm d} \propto (16 J^2-3\alpha_R^2
k_F^2-9\alpha_R^2 k_F^2 \cos(2\theta))$. Therefore, the Fermi sea
contribution induces an intrinsic angular dependence in $\tau_{\rm
  d}$, which increases with $\frac{|\alpha_R| k_F}{J}$ irrespective of
the Fermi surface distortion. The results in
Fig. \ref{Fig3}, which are obtained numerically, include the Fermi
surface distortion, so that the nontrivial angular dependence of
$\tau_{\rm d}$ results from the combined effects of the intrinsic
Fermi sea contribution and the Fermi surface distortion.  For example,
Fig. \ref{Fig3}(a) shows a sharp difference in the angular dependence
of $\tau_{\rm d}$ for $r>1$ and $r<1$. This is qualitatively similar
to the results of $\tau_{\rm f}$ shown in Fig. \ref{Fig2}(b),
showing that the Fermi surface distortion also has a role in the
angular dependence of $\tau_{\rm d}$.

The sign of $\tau_{\rm d}$ does not change with the magnetization
angle despite the strong angular dependence, similar to the behavior
of $\tau_{\rm f}$. When $\theta=\pi/2$ (Fig. \ref{Fig3}(d)), a steep
increase of $\tau_{\rm d}$ is obtained at $\phi=\pi/4$ and $3\pi/4$,
originating from the shape of the Fermi surface. We expect that this
strong dependence of $\tau_{\rm d}$ on $\phi$ can be observed in
epitaxial bilayers but may be absent in sputtered bilayers as
sputtered thin films consist of small grains with different lattice
orientation in the film plane. However, the dependence of $\tau_{\rm
  d}$ on the polar angle $\theta$ (Fig. \ref{Fig3}(a) and (c)) is
irrelevant to this in-plane crystallographic issue so we expect that
it will be observable in experiments when the interfacial spin-orbit
coupling is comparable to or stronger than the exchange coupling. We
note that a strong dependence of $\tau_{\rm d}$ on $\theta$ (but a
very weak dependence on $\phi$) was experimentally observed in
sputtered bilayers.~\cite{Garello:2013}

We finally illustrate the connection between $\tau_{\rm d}$
(i.e. ${\bf T}_{(2)}$) and the Berry phase.  This examination is
motivated by Ref.~\onlinecite{Kurebayashi:2014}, which called ${\bf
  T}_{(2)}$ the Berry phase contribution.  To clarify the connection,
it is useful to express the Fermi sea contribution ${\bf s}_{(2)}$ in
the Kubo formula form,
\begin{widetext}
  \begin{equation}
    {\bf s}_{(2)}=\frac{1}{2}eE\hbar^2 A {\rm Im}\sum_{ab}\int \frac{d^2 k}{(2\pi)^2}
    \left[ f_a({\bf k})-f_b({\bf k}) \right]
    \frac{ \langle {\bf k},a | {\bm \sigma} |{\bf k},b\rangle \langle {\bf k},b| v_x | {\bf k},a \rangle }{\left[ E_a({\bf k})-E_b({\bf k})+2i\delta\right]^2},
    \label{eq:Kubo-formulat-for-s2}
  \end{equation}
\end{widetext}
where $a$, $b$ are band indices, and $\delta$ is an infinitesimally
small positive constant.  In the present case, $|{\bf k},a\rangle$ is
either $|\psi_{{\bf k},+}\rangle$ or $|\psi_{{\bf k},-}\rangle$.  One
then uses the relations
\begin{equation}
  v_x=\frac{1}{\hbar} \frac{\partial H_0({\bf k},{\bf M})}{\partial k_x}, \ \
  \sigma_\alpha=\frac{1}{J}\frac{\partial H_0({\bf k},{\bf M})}{\partial M_\alpha},
  \label{eq:Operator-relations}
\end{equation}
where the notation $H_0({\bf k},{\bf M})$ emphasizes that the
unperturbed Hamiltonian $H_0$ is a function of the momentum ${\bf k}$
and the magnetization ${\bf M}$.  Note that here we use ${\bf M}$
instead of $\hat{\bf M}$ since one needs to relax the constraint
$|\hat{\bf M}|=1$ to establish the connection with the Berry phase.
Equation~(\ref{eq:Operator-relations}) allows one to convert the
numerator of Eq.~(\ref{eq:Kubo-formulat-for-s2}) as follows,
\begin{widetext}
  \begin{equation}
    \langle {\bf k},a | {\bm \sigma} |{\bf k},b\rangle=-\frac{1}{J}\left[ E_a({\bf k})-E_b({\bf k}) \right] \langle {\bf k},a | \nabla_{\bf M} |{\bf k},b \rangle, \ \
    \langle {\bf k},b| v_x | {\bf k},a \rangle=-\frac{1}{\hbar}\left[ E_b({\bf k})-E_a({\bf k}) \right]  \langle {\bf k},b | \frac{\partial}{\partial k_x} |{\bf k},a \rangle.
  \end{equation}
\end{widetext}
Thus the numerator of Eq.~(\ref{eq:Kubo-formulat-for-s2}) acquires the
factor $\left[ E_a({\bf k})-E_b({\bf k}) \right]^2$, which cancels the
denominator in the limit $\delta\rightarrow 0$.  Then one of the two
summations for the band indices $a$ and $b$ can be performed to
produce
\begin{widetext}
  \begin{equation}
    [{\bf s}_{(2)}]_\alpha=\frac{1}{2}\frac{eE\hbar A}{J} \sum_{a}\int \frac{d^2 k}{(2\pi)^2}
    f_a({\bf k})
    \left[
      \frac{\partial}{\partial k_x} {\cal A}_{M_\alpha}^{a}({\bf k})
      -\frac{\partial}{\partial M_\alpha} {\cal A}_{k_x}^{a}({\bf k})
    \right],
    \label{eq:Kubo-formulat-for-s2-Berry-phase}
  \end{equation}
\end{widetext}
where the spin-space Berry phase ${\cal A}_{M_\alpha}^{a}({\bf k})$
and the momentum-space Berry phase ${\cal A}_{k_x}^{a}({\bf k})$ are
defined by
\begin{eqnarray}
  {\cal A}_{M_\alpha}^{a}({\bf k})&=&i \langle {\bf k},a | \frac{\partial}{\partial M_\alpha} | {\bf k},a \rangle \\ \nonumber
  {\cal A}_{k_x}^{a}({\bf k})&=&i \langle {\bf k},a | \frac{\partial}{\partial k_x} | {\bf k},a \rangle.
\end{eqnarray}
Here these Berry phases are manifestly real.
Equation~(\ref{eq:Kubo-formulat-for-s2-Berry-phase}) establishes the
connection between ${\bf T}_{(2)}$ and the spin-momentum-space Berry
phase.

A few remarks are in order.  First, through an explicit
evaluation of the Berry phases, one can verify that the integrand of
Eq.~(\ref{eq:Kubo-formulat-for-s2-Berry-phase}) generates ${\bf
  s}_{{\bf k}(2)}^\pm$ in Eq.~(\ref{Eq:skfem2}) precisely.  Second,
Eq.~(\ref{eq:Kubo-formulat-for-s2-Berry-phase}) contains the
occupation function $f_a({\bf k})$ itself rather than difference
between the occupation functions or derivatives of the occupation
function.  Thus ${\bf s}_{(2)}$ may be classified as a Fermi sea
contribution.  We note, however, that the
Fermi sea contribution Eq.~(\ref{eq:Kubo-formulat-for-s2-Berry-phase}) may be
converted to a different form~\cite{Haldane:2004}, where the net
contribution is evaluated only at the Fermi surface. To demonstrate
this point, we integrate
Eq.~(\ref{eq:Kubo-formulat-for-s2-Berry-phase}) by parts, which generates
\begin{widetext}
  \begin{equation}
    [{\bf s}_{(2)}]_\alpha=\frac{1}{2}\frac{eE\hbar A}{J} \sum_{a} \int \frac{d^2 k}{(2\pi)^2}
    \left[
      -\frac{\partial f_a({\bf k})}{\partial k_x}
      {\cal A}_{M_\alpha}^{a}({\bf k})
      + \frac{\partial f_a({\bf k})}{\partial M_\alpha}
      {\cal A}_{k_x}^{a}({\bf k})
    \right].
  \end{equation}
\end{widetext}
Note that in the zero temperature limit, both $\partial f_a({\bf
  k})/\partial k_x$ and $\partial f_a({\bf k})/\partial M_\alpha$ are
nonzero only at the Fermi surface, and thus the net contribution to
${\bf s}_{(2)}$ depends only on properties evaluated at the Fermi
surface.  In this sense, this Fermi surface contribution is analogous
to Friedel oscillations.  Friedel oscillations form near surfaces
when electrons reflect and the incoming and
outgoing waves interfere.  Then, each electron below the Fermi energy
makes an oscillatory contribution to the density with a wavelength
that depends on the energy.  However, integrating up from the bottom
of the band to the Fermi energy gives a result that only depends on
the properties of the electrons at the Fermi energy where there is a
sharp cut-off in the integration.

\section{Summary}
\label{sec:sum}

We use simple models to examine the angular dependence of spin-orbit
torques as a function of the ratio of the spin-orbit interaction to
the exchange interaction.  We find that both the field-like and
damping-like torques are angle independent when the spin-orbit
coupling is weak but become angle-dependent when the spin-orbit
coupling becomes comparable to the exchange coupling.  When the
spin-orbit coupling becomes much stronger than the exchange coupling,
the angular dependence of the field-like torque goes away, but that of
the damping-like torque remains.  The angular dependence of the
field-like torque becomes significant when the spin-orbit coupling
becomes strong enough to distort the Fermi surface so that it changes
when the direction of the magnetization changes.  On the other hand,
the angular dependence of the damping-like torque is caused by the
combined effects of the intrinsic Fermi sea contribution and the Fermi
surface distortion.  We expect that these qualitative conclusions will
hold for more realistic treatments of the interface.  The strong
angular dependence of the spin-orbit torques will significantly impact
their role in large amplitude magnetization dynamics like switching or
domain wall motion.  This suggests caution when comparing measurement
of the strength of torques with the magnetizations in different
directions.

\acknowledgments

K.-J.L. acknowledges support from the NRF (2011-028163,
NRF-2013R1A2A2A01013188) and under the Cooperative Research Agreement
between the University of Maryland and the National Institute of
Standards and Technology Center for Nanoscale Science and Technology,
Award 70NANB10H193, through the University of Maryland. H.-W.L. was
supported by NRF (2013R1A2A2A05006237) and MOTIE (Grant
No. 10044723). A.M. acknowledges support by the King Abdullah
University of Science and Technology. D.G. acknowledges support from
the Global Ph.D. Fellowship Program funded by NRF (2014H1A2A101).

\appendix
\section{Derivation of Eq.~(9)}
\label{sec:appendixA}

Here, we derive the Fermi sea contribution of the spin-orbit
torque. First, we calculate the change of the eigenstates in the
presence of an external electric field. Second, we calculate the
resulting spin accumulation. We use time-dependent perturbation
theory, adiabatically turning on the electric field, which is treated
as the perturbation.  We adopt time-dependent perturbation approach
instead of the Kubo formula for pedagogical reasons since it directly
shows how the states change due to the perturbation. One can show that
both approaches give the same result.

Let us consider an in-plane electric field
$\mathbf{E'}(t)=\mathbf{E}\exp{(\delta t)}$ , where $\exp{(\delta t)}$
gives the adiabatic turning-on process. The
electric field starts to increase from $t=-\infty$ until $t=0$, for
very small $\delta$ which will be set to be zero at the end of the
calculation. This is represented by the vector potential
$\mathbf{A}=-t\exp{(\delta t)}\mathbf{E}$ since $\mathbf{E}=-\partial
\mathbf{A}/{\partial t}$. In the presence of a vector potential, the momentum
operator $\mathbf{p}$ is replaced by $\mathbf{p}+e\mathbf{A}$. Thus,
the total Hamiltonian becomes
\begin{align}
  H&=\frac{\left(\mathbf{p}+e\mathbf{A} \right)^2}{2m}+\frac{\alpha_R}{\hbar}\boldsymbol{\sigma}\cdot \left[ \left( \mathbf{p}+e\mathbf{A}\right)\times \hat{\mathbf{z}} \right] + J\boldsymbol{\sigma}\cdot \hat{\mathbf{M}} \nonumber
  \\
  &=H_0+H_1(t)+ \mathcal{O}(\mathbf{E}^2)
\end{align}
where
\begin{align}
  H_1(t)=&- \frac{e\left( \mathbf{E}\cdot\mathbf{p} \right)}{m}t\exp{(\delta t)}
  \\
  &+\frac{\alpha_R}{\hbar} [(e\mathbf{E}\times \boldsymbol{\sigma}) \cdot \hat{\mathbf{z}}] t\exp{(\delta t)}. \nonumber
\end{align}
Here, the first term comes from the kinetic energy and the the second
term from the Rashba spin-orbit coupling. In the interaction
picture, the propagator of the order of $\mathcal{O}(\mathbf{E}^1)$ is
\begin{equation}
  U^{\textup{(I)}}_1=-\frac{i}{\hbar}\int_{-\infty}^{0}dt\ \mathcal{H}^{\textup{(I)}}_1(t)
\end{equation}
where
\begin{align}
  \mathcal{H}^{\textup{(I)}}_1(t)
  =&e^{i\mathcal{H}_0t/\hbar}\mathcal{H}_1(t) e^{-i\mathcal{H}_0t/\hbar} \nonumber
  \\
  =&- \frac{e\left( \mathbf{E}\cdot\mathbf{p} \right)}{m}t\exp{(\delta t)} \nonumber
  \\
  &+  \frac{\alpha_R}{\hbar} [(e\mathbf{E}\times \boldsymbol{\sigma}^{\textup{(I)}}(t)) \cdot \hat{\mathbf{z}}] t\exp{(\delta t)} ,
\end{align}
and
\begin{align}
  \boldsymbol{\sigma}^{\textup{(I)}}(t)=&e^{i\mathcal{H}_0t/\hbar}\boldsymbol{\sigma} e^{-i\mathcal{H}_0t/\hbar} \nonumber
  \\
  =&\boldsymbol{\sigma}\cos\left ( \frac{2\epsilon_k t}{\hbar} \right )+(\hat{\mathbf{n}}\times\boldsymbol{\sigma})\sin\left ( \frac{2\epsilon_k t}{\hbar} \right ) \nonumber
  \\
  &+\hat{\mathbf{n}}(\hat{\mathbf{n}}\cdot\boldsymbol{\sigma})\left [ 1- \cos\left ( \frac{2\epsilon_k t}{\hbar} \right )\right ].
\end{align}
Here we define
\begin{align}
  \epsilon_\mathbf{k} &=\left | J\hat{\mathbf{M}}+\alpha_R\mathbf{k}\times\hat{\mathbf{z}} \right |,
\end{align}
and
\begin{align}
  \hat{\mathbf{n}}&=\frac{1}{\epsilon_\mathbf{k}}\left ( J\hat{\mathbf{M}}+\alpha_R\mathbf{k}\times\hat{\mathbf{z}} \right ).
\end{align}
Thus,
\begin{equation}
  U^{\textup{(I)}}_1(\mathbf{k})=U^{\textup{(I)}}_{1(a)}(\mathbf{k})+U^{\textup{(I)}}_{1(b)}(\mathbf{k}),
\end{equation}
where
\begin{equation}
  U^{\textup{(I)}}_{1(a)}(\mathbf{k})=-i\frac{e\left( \mathbf{E}\cdot\mathbf{k} \right)}{m}\frac{1}{\delta^2},
\end{equation}
and
\begin{widetext}
  \begin{align}
    U^{\textup{(I)}}_{1(b)}(\mathbf{k})=-\frac{i}{\hbar}\frac{\alpha_R}{\hbar} & \left \{
      \frac{1}{2}[(e\mathbf{E}\times \boldsymbol{\sigma}) \cdot \hat{\mathbf{z}}] \left (
        \frac{1}{(2\epsilon_\mathbf{k}/\hbar + i\delta)^2}
        +
        \frac{1}{(2\epsilon_\mathbf{k}/\hbar - i\delta)^2}
      \right ) \right.
    \nonumber
    \\
    &+
    \frac{1}{2i}[(e\mathbf{E}\times (\hat{\mathbf{n}}\times \boldsymbol{\sigma}))\cdot \hat{\mathbf{z}}]
    \left (
      \frac{1}{(2\epsilon_\mathbf{k}/\hbar + i\delta)^2}
      -
      \frac{1}{(2\epsilon_\mathbf{k}/\hbar - i\delta)^2}
    \right ) \nonumber
    \\
    &\left. +
      [(e\mathbf{E}\times \hat{\mathbf{n}})\cdot \hat{\mathbf{z}}](\hat{\mathbf{n}}\cdot \boldsymbol{\sigma})
      \left [
        -\frac{1}{\delta^2}
        -
        \frac{1}{2}\left (
          \frac{1}{(2\epsilon_\mathbf{k}/\hbar + i\delta)^2}
          +
          \frac{1}{(2\epsilon_\mathbf{k}/\hbar - i\delta)^2}
        \right )
      \right ]
    \right \}.
  \end{align}
\end{widetext}
Thus, the change in the state due to the adiabatically turned on
electric field is given by
\begin{equation}
  \delta \ket{\psi_{\mathbf{k},\pm}} = U^{\textup{(I)}}_{1}(\mathbf{k})\ket{\psi_{\mathbf{k},\pm}}.
\end{equation}

Now, the spin accumulation arising from the changes in the occupied states is
\begin{eqnarray}
  s^\pm_{\mathbf{k}(2)}
  &=&\frac{\hbar}{2}\left[ 
   \left( \delta \bra{\psi_{\mathbf{k},\pm}} \right)
    \boldsymbol{\sigma}
     \ket{\psi_{\mathbf{k},\pm}}  
    +
      \bra{\psi_{\mathbf{k},\pm}} 
    \boldsymbol{\sigma}
    \left(  \delta \ket{\psi_{\mathbf{k},\pm}} \right)
    \right]_{\delta\rightarrow 0} \nonumber
  \\
  &=&
  \frac{\hbar}{2}\times 2\textup{Re}\left [ \bra{\psi_{\mathbf{k},\pm}}\boldsymbol{\sigma}U^{(1)}_1(\mathbf{k})\ket{\psi_{\mathbf{k},\pm}} \right ]_{\delta\rightarrow 0} \nonumber
  \\
  &=& \pm\frac{\hbar}{2}\alpha_R e E \left\{
    \frac{ J}{2\epsilon_\mathbf{k}^3}[\hat{\mathbf{M}}\times (\hat{\mathbf{z}}\times \hat{\mathbf{E}})]+\frac{\alpha_R}{2\epsilon_\mathbf{k}^3}(\hat{\mathbf{E}}\times \mathbf{k})\right\}, \nonumber
  \\
\end{eqnarray}
where $\pm$ indicates majority/minority bands, respectively. When the
electric field is applied along the $\hat{\mathbf{x}}$ direction, we
arrive at Eq. (6). Note that $U^{\textup{(I)}}_{1(a)}$ does not
contribute to the spin expectation value since it is purely imaginary.

($**$) These authors equally contributed to this work.

\end{document}